\documentclass[aoas,dvips]{arximspdf}
\usepackage{mathrsfs}

 \referstodoi{10.1214/09-AOAS312}
\doi{10.1214/09-AOAS312REJ}
\volume{3}
\issue{4}
\pubyear{2009}
\firstpage{1303}
\lastpage{1308}

\makeatletter
\DeclareMathAlphabet\mathcaligr{OMS}{cmsy}{m}{n}
\renewcommand{\mathcal}{\mathcaligr}
\newproclaim{definition}{Definition}
\makeatother

\begin{document}
\begin{frontmatter}

\title{Rejoinder: Brownian distance covariance}
\runtitle{Rejoinder}

\begin{aug}
\author[A]{\fnms{G\'{a}bor J.} \snm{Sz\'{e}kely}\ead[label=e2]{gszekely@nsf.gov}}
\and
\author[B]{\fnms{Maria L.} \snm{Rizzo}\ead[label=e1]{mrizzo@bgsu.edu}\corref{}}
\runauthor{G. J. Sz\'{e}kely and M. L. Rizzo}
\pdfauthor{Gabor J. Szekely, Maria L. Rizzo}
\affiliation{Bowling Green State University, Hungarian Academy of
Sciences and\break
Bowling Green State University}

\address[A]{Department of Mathematics and Statistics\\
Bowling Green State University\\
Bowling Green, Ohio 43403\\
USA\\
and\\
R\'{e}nyi Institute of Mathematics\\
Hungarian Academy of Sciences\\
Hungary\\
\printead{e2}}

\address[B]{Department of Mathematics and Statistics\\
Bowling Green State University\\
Bowling Green, Ohio 43403\\
USA\\
\printead{e1}}
\end{aug}




\end{frontmatter}

First of all we want to thank the editor, Michael Newton, for
leading the review and discussion of our work.

We also want to thank all discussants for their interesting comments.
Some of them are in fact short research papers that expand the scope of
Brownian Distance Covariance. Many of the comments emphasized the
existence of some competing notions like maximal correlation; others
requested further clarifications or suggested several extensions. Most
of the comments were theoretical in nature.
We do hope that once our new correlation is applied in practice we shall
receive comments from the broader community of applied statisticians.
Let us now continue with replies to the discussions collectively by
grouping the topics.

\section{Unbiased distance covariance}

In the discussion Cope observes that the distance dependence statistics
are biased, and that this bias may be substantial and increasing with
dimension. As he points out, in genomic studies, high dimension and
small sample sizes are common.

In this section we present an unbiased estimator of the population
distance covariance, define a corrected distance correlation statistic
$C_n$, and propose a simple decision rule for the high dimension, small
sample size situation.

The expected value of $\mathcal V_n^2$ is
$E[\mathcal V^2_n (\mathbf X, \mathbf Y) ] = \frac{n-1}{n^2} [
(n-2)\mathcal V^2(X,Y) + \mu_1 \mu_2  ],$ where $\mu
_1=E|X-X'|$ and $\mu_2=E|Y-Y'|$. An unbiased estimator of $\mathcal
V^2(X,Y)$ can be defined as follows.
\begin{definition}
\[
 {U}_n (\mathbf X, \mathbf Y) =
\frac{n^2}{(n-1)(n-2)} \biggl[\mathcal V_n^2(\mathbf X, \mathbf Y) -
\frac{T_2}{n-1} \biggr], \qquad n \geq3,
\]
where $T_2$ is the statistic defined in Theorem 1.
\end{definition}

We proposed to normalize the $V$-statistic $n \mathcal V_n^2$ by
dividing by $T_2$. Under independence, it follows from Corollary 2(i) that
\[
\frac{n  {U}_n}{T_2} =
\frac{n^2}{(n-1)(n-2)}  \biggl[\frac{n\mathcal V_n^2}{T_2} - \frac
{n}{n-1} \biggr]
\stackrel {\mathscr D} \longrightarrow\sum\limits_{k=0}^\infty
\lambda_k
(Z_k^2 -1) \quad\mbox{as } n \to\infty,
\]
which is the limiting distribution of the corresponding $U$-statistic.

A modified distance correlation statistic $C_n$ can be defined by
substituting in the original definition of $\mathcal R^2_n$ the
unbiased estimators $U_n$. It can be shown that $U_n (\mathbf X,\mathbf
X) \geq0$ for $n \geq3$, so that ${U_n(\mathbf X) U_n(\mathbf Y)} >
0$ whenever $ {\mathcal V}_n^2(\mathbf X) {\mathcal V}_n^2(\mathbf Y) >
0$, $n \geq3$.
\begin{definition} The corrected distance correlation for sample sizes
$n \geq3$ is
\[
C_n(\mathbf X,\mathbf Y) =
\cases{\displaystyle
\frac{U_n(\mathbf X,\mathbf Y)}
{\sqrt{U_n(\mathbf X)
U_n(\mathbf Y)}}, &\quad $U_n(\mathbf X)
U_n(\mathbf Y) > 0$; \cr
0, &\quad  otherwise.
}
\]
If $n=1$ or $n=2$ define $C_n=1$.
\end{definition}

If $X$ and $Y$ are independent, $(p+q)/n$ is large and $n$ is
moderately large, one can compare $n C_n$ with percentiles of a
Normal($0, \sigma^2=2)$ distribution, under very general conditions on
the distributions of $X$ and $Y$.

\section{Other measures of dependence, old and new}

Bickel and Xu mentioned canonical correlation $\rho$, rank correlation
$r$ and R\'{e}nyi correlation~$R$. Of these, only~$R$ is the one which
vanishes if and only if $X$ and $Y$ are independent.
A big advantage of dCor vs $R$ is that dCor is much easier to compute.
In the discussion there is a method to approximate R\'{e}nyi's $R$, but
frankly we do not think that the simplicity of computing or even
approximating $R$ is comparable to the simplicity of computing  Pearson
correlation. Part of the reason is that there is no explicit
formula for computing $R$ in general. On the other hand, we have an
explicit formula to compute dCov, and practitioners or applied
statisticians should find it easy to use.

For the first named author it was heartwarming to see several references
to R\'{e}nyi because R\'{e}nyi was his first advisor and mentor. In his 1959
paper, R\'{e}nyi \cite{renyi59} characterized $R$ with seven ``natural
postulates.''
His last postulate is that the dependence measure equals the absolute
value of  Pearson correlation for bivariate normal distributions.
This axiom does not hold in our case, although dCor is a deterministic
function of  Pearson correlation. It would be nice to extend R\'{e}nyi's
theorem and prove a joint characterization of $R$ and dCor.

Bickel and Xu remind us that ``if $R=1$ then then there exist nontrivial
functions~$f$ and $g$ such that $P(f(X)=g(Y))=1 \ldots.$''
However, the following example suggests that this is not necessarily a
desirable property. Consider random variables $X=\sin kU$ and $Y=\sin
mU$, where $U$ is uniformly distributed on $(0, 2\pi)$, and $k,m$ are distinct
positive integers. Their Pearson correlation is 0, yet for Chebyshev
polynomials $\{T_k\}$, we have
\begin{eqnarray*}
T_k( \cos 2mU)&=&T_m(\cos 2kU) \\
&=& T_k(1-2Y^2) = T_m(1-2X^2).
\end{eqnarray*}
Thus, $R=1$ even though in many cases $X$ and $Y$ are heuristically quite
unrelated: neither $f$ nor $g$ is invertible, $Y$ is not a function
of $X$ and vice versa; exceptions are when $m$ is an odd multiple of $k$.
Our simulations suggest that $0 < \mathrm{dCor} < 1/3$ for the examples
above, and reaches its maximum when $m=3k$.

Because $X$ and $Y$ are not
independent, it is not surprising that the CLT does not hold for
\[
S_n = \sin U + \sin2U + \cdots+ \sin nU.
\]
Nevertheless, it can be surprising that $S_n$ tends to C$/$2 in
distribution, as $n\to\infty$, where C is a standard Cauchy random
variable.
(It is not a misprint that we did not divide by $\sqrt{n}$; here we do
not need any kind of normalization.) For the proof of this result and
generalizations to other ``trigonometric coins,'' other orthogonal
series, and finite Fourier series, see ``Trigonometric Coins''
\cite{SMPZ}.
The general infinite Fourier series case is an open problem. One of the
advantages of dCov is that in terms of $\mathrm{dCov}=0$ type conditions
we can prove general CLTs for strongly stationary series (Sz\'{e}kely and
Bakirov \cite{TR08-01}).

Further dependence measures can be found in the discussion of Gretton,
Fukuzimu and Sriperumbudur. We recognize the theoretical importance of
RKHS-based dependence measures, but they do not look as simple as our
distance covariance, and they do not seem to be formal extensions of
Brownian distance covariance because our weight function (2.4) is not
integrable.

\section{Generalizations to metric spaces}

One can easily extend the definition of Brownian distance covariance via
formula (2.8) to all metric spaces; all we need is to replace the
Euclidean distances between observations with their metric distances.
Thus in principle we can measure the dependence between two samples
where the sample elements come from two arbitrary metric spaces.
In order to prove counterparts of our theorems, we need further
restrictions. One of the possible approaches is to try to represent the
abstract samples in finite dimensional Euclidean spaces such that the
distances $a_{kl}, b_{kl}$ become interpoint distances in these Euclidean
spaces. Necessary and sufficient conditions are established in the
multidimensional scaling literature (see, e.g., Mardia, Kent and Bibby
\cite{mkb79}, Chapter~14). When such a representation is possible, many
theorems in this paper can be extended to measuring and testing
independence of random vectors that take values in abstract metric
spaces. For example, the metric space extension is applicable for
testing independence of categorical data. They are not in Euclidean
spaces, but their association can be used as a distance.

A very important area of applications is how to measure the
dependence of stochastic processes. In this respect, infinite
dimensional extensions of our paper are crucial, so we commend the
discussion of Kosorok. Because of his work we now have an extension of
our theorems to certain Hilbert spaces.

\section{Invariance}

Our test statistic is scale invariant and also rotation invariant.
Cram\'{e}r--von Mises type test statistics, mentioned, for example, in
R\'{e}millard's discussion Section 2, are not rotation invariant. This is
a major problem if one wants to extend the measure to metric spaces. Let
us emphasize that our test \textit{procedure} is invariant with respect to
marginal distributions, even though the test statistic is not. On the
other hand, it is true that we can easily make our dependence measure
even more invariant (invariant with respect to the marginals and with
respect to monotone transformations) if we apply the transformations
suggested in Section 1 of R\'{e}millard. The negative side of this is that
we might lose power, especially if the sample size is small.

R\'{e}millard asked if certain dependence measures can be written in our
form.
The general answer is no, because the well-known measures such as
Kendall's tau and many other rank based measures do not characterize
independence, or the statistics are not rotation invariant (e.g., Cram\'{e}r--von Mises), or like maximal correlation they do not have an
explicit computing formula, or may not be defined for arbitrary
dimension (e.g., Feuerverger's measure \cite{feuer93}).

Invariance with respect to monotone transformations in one dimension
suggests rank type tests such as Feuerverger \cite{feuer93}, but they
have the disadvantage of being one-dimensional. We can also eliminate
all kinds of moment
conditions by transforming $X$ and $Y$ to bounded random variables first
and then compute their distance covariance, but then there is an
arbitrariness
in choosing these bounded functions. In one dimension the rank is a
natural choice.
Section 2 of R\'{e}millard's discussion proposes a natural rank based
transformation
for the multivariate case.

\section{Applications}

Genovese asks about the generality or required conditions for the test
of nonlinearity, Example 6. The application of dCov to testing for
nonlinearity requires only that the linear model $Y=X\beta+\varepsilon$
can be estimated, and that observations ($\mathbf X, \hat\varepsilon
)$ are i.i.d.
The existence of first moments is implicit in the linear model
specification. Distance covariance is defined in arbitrary dimension,
so the procedure can be applied to models with a multivariate response.
This expands the scope of the test, because models can often be
specified with a multivariate response and i.i.d. errors.

The extension of distance covariance methods to non-i.i.d. samples would be
very important for applications; see, e.g., R\'{e}millard's discussion
Section 3 on
the application to time series: \textit{Serial Brownian Distance
Correlation}. We
agree with R\'{e}millard that ``there are still many interesting
avenues'' to
explore in this context.

\section{Simplicity/complexity}

Our formula (2.8) to compute dCov is not only simple, it has an obvious
formal similarity to  Pearson product moment covariance,
except that we need to average $n^2$ products.
Genovese comments that the $O(n^2)$ computational complexity
of $\mathcal R_n$ or $\mathcal V_n$ can be burdensome for very
large~$n$. However, the simplicity of the computing formula (2.8)
in terms of products $A_{kl}B_{kl}$ provides economies of reusable
computations. The distances need only be computed once in the
permutation test implementation, as the permutation of sample
indices of $\mathbf Y$ corresponds to permutations of indices of
$B_{kl}$, for
example.

If we compare the complexity of our statistic (2.8) to the
complexity of other measures of dependence (including, e.g., RKHS-based
methods
suggested by the discussants Gretton et al.,
or our own measure proposed in Bakirov, Rizzo and Sz\'{e}kely
\cite{brs06}), then the superiority of Brownian distance covariance is
clear. On top of that, one can compute dCov even if the $X$ sample and
the $Y$ sample are in completely different metric spaces, because it is
not necessary to add or multiply the sample elements; we need only
operations on their real valued distances. This is a significant
advantage if we want to measure the dependence of apples and oranges,
even infinite dimensional ones.

\section{Distance covariance vs product-moment covariance and how to teach them}

After noticing that  Pearson and distance covariance are two
different special cases of a general notion of covariance with respect
to stochastic processes, we have not
explored the boundaries of this generalization. We focused on the two
most natural and simplest cases: Brownian covariance and Pearson
covariance. Feuerverger raises some interesting questions in this
direction at the end of his discussion. R\'{e}millard also raises some
questions on the role of stochastic
processes~$U, V$. Genovese's discussion sheds some light on these
questions. Although we have not yet explored the frontiers of these
extensions, these questions and the
research of Genovese on this topic are indeed interesting.

For more than a century  Pearson correlation has dominated the world of
measuring dependence. Even though we know that for nonnormal
distributions, product-moment correlation does not characterize
independence (does not really measure what we want) for reasons of
simplicity, perhaps, it is the first and sometimes the only measure of
dependence that students may see. Here Genovese raises a good
pedagogical question: should distance correlation be introduced in our
teaching at an introductory level?
Indeed, we agree that the idea of distance correlation is
understandable even at the undergraduate level (without proofs), and
one could then continue with product-moment correlation for normal
distributions obtained with exponent $\alpha=2$.

\section{Final comments}

Our test of independence is implemented in R as part of the ``energy''
package \cite{R,energy}. The explanation of this cover name is that
Newton's potential
energy is a function of the Euclidean distances between objects in a
gravitational space. In \textit{energy statistics} the ``objects'' are the
elements of the statistical
sample, and the statistics are functions of the Euclidean distances
between the sample elements. These statistics, the statistical
potential energies, govern the cosmos of our paper.

\printaddresses


\begin{thebibliography}{9}

\bibitem{brs06}
 \textsc{Bakirov, N. K.,  Rizzo, M. L.} and \textsc{Sz\'{e}kely, G. J.} (2006). A multivariate
nonparametric test of independence. \textit{J. Multivariate Anal.}
\textbf{93} 1742--1756.
\MR{2298886}

\bibitem{feuer93}
\textsc{Feuerverger, A.} (1993). A consistent test for bivariate dependence.
\textit{Int. Statist. Rev.} \textbf{61} 419--433.


\bibitem{mkb79}
\textsc{Mardia, K. V.}, \textsc{Kent, J. T.} and \textsc{Bibby, J. M.} (1979).
\textit{Multivariate Analysis}. Academic Press, London.
\MR{0560319}

\bibitem{R}
\textsc{R Development Core Team} (2009). R: A language and environment for
statistical computing. R Foundation for Statistical Computing,
Vienna, Austria. ISBN 3-900051-07-0. Available at \href{http://www.R-project.org}{http://www.R-project.org}.

\bibitem{renyi59}
\textsc{R\'{e}nyi, A.} (1959). On measures of dependence. \textit{Acta Math.
Acad. Sci. Hungary} \textbf{10} 441--451.
\MR{0115203}

\bibitem{energy}
 \textsc{Rizzo, M. L.} and \textsc{Sz\'{e}kely, G. J.} (2008). energy: {E}-statistics
(energy statistics). R package version 1.1-0. Available at
\href{http://cran.case.edu/web/packages/energy/}{http://cran.case.edu/web/packages/energy/}.

\bibitem{TR08-01}
 \textsc{Sz\'{e}kely, G. J.} and  \textsc{Bakirov, N. K.} (2008).
 Brownian covariance and
{CLT} for stationary sequences. Technical Report No. 08-01.
Dept.  Mathematics \& Statistics,
 Bowling Green State University.

\bibitem{SMPZ}
\textsc{Sz\'{e}kely, G. J., M\'{o}ri, T. F., Phadke, V.} and \textsc{Zirbel, C.} (2008).
Trigonometric coins. Unpublished manuscript.


\end{thebibliography}
\end{document}